\documentstyle[epsf,preprint,aps,eqsecnum,amsfonts,tighten]{revtex}
\newcommand{\beq}{\begin{equation}}
\newcommand{\eeq}{\end{equation}}
\begin{document}

\def\bold#1{\setbox0=\hbox{$#1$}%
     \kern-.025em\copy0\kern-\wd0
     \kern.05em\copy0\kern-\wd0
     \kern-.025em\raise.0433em\box0 }
{\tighten
\preprint{\vbox{\hbox{CERN-TH/95-182}
                \hbox{WIS-95/27/Jul-PH}
                \hbox{SCIPP-95/33}
                \hbox{hep-ph/9507213}
                \hbox{\phantom{July, 1995}}
                \hbox{\phantom{July, 1995}}
}}
\title{QCD corrections to charged Higgs-mediated
       $\bold{b \to c \, \tau\,\nu}$ decay}
\author{Yuval Grossman,$^a$ Howard E. Haber\footnote{Permanent
address: Santa Cruz Institute for Particle Physics, University of
California, Santa Cruz, CA 95064 USA.}$^{,b}$  and  Yosef Nir $^a$}
\address{ \vbox{\vskip 0.5truecm}
  $^a$Department of Particle Physics \\
      Weizmann Institute of Science, Rehovot 76100, Israel \\
\vbox{\vskip 0.1truecm}
  $^b$CERN, TH-Division \\
  CH-1211 Geneva 23, Switzerland}
\maketitle
\begin{abstract}%
We calculate the $O(\alpha_s)$ corrections to the
charged Higgs mediated inclusive semi-tauonic $B$ decay.
By working in the Landau gauge, we demonstrate how
to obtain the charged Higgs contributions (both direct
and interference terms) from the known QCD corrections
to $t \to b H^+$. Combining
our results with $O(1/m_b^2)$ corrections to the spectator model
and comparing the theoretical prediction with the recent experimental
measurements, we find a model-independent
$2\sigma$ upper bound on the ratio of Higgs vacuum expectation values,
$\tan\beta<0.52\,(m_H/1\,{\rm GeV})$.

\end{abstract}
\vfill
\vbox{\hbox{CERN-TH/95-182}
       \hbox{July, 1995}}
}
\newpage

\def\tgb{\tan\beta}
\def\tgbs{\tan^2\beta}
\def\ctgb{\cot\beta}
\def\ctgbs{\cot^2\beta}
\def\al{\alpha_s}
\def\bctn{B \to X \, \tau\,\nu}
\def\bcen{B \to X \, e\,\nu}
\def\bcln{B \to X \, \ell \,\nu}
\def\G{\Gamma}
\def\le{\left}
\def\ri{\right}
\def\M{{\cal M}}
\def\Lra{\Longrightarrow}
\def\r{\rho}
\def\e{\rho_\tau}
\def\eps{\epsilon}

\narrowtext

\section{Introduction}

The $\bctn$ decay is sensitive to
extensions of the standard model (SM) Higgs sector.  In particular,
for two Higgs doublet models (2HDM) \cite{hhg}, the experimental
measurement of this decay provides the strongest upper bound on
$\tgb/m_H$, when the ratio of Higgs vacuum expectation values,
$\tan\beta$, is large \cite{KrPo,Kali,YZ}.
Since the charged Higgs boson contributes at tree level, its contribution
cannot be canceled by other new particles in the theory.
Consequently, this bound holds in all SU(2)$\times$U(1) electroweak
models with a charged Higgs boson, independently of other details
of the model. It is therefore important to calculate the charged
Higgs contribution with high accuracy.

The leading order calculation is based on tree diagrams of the
spectator model. Corrections to spectator model results are given by the
$1/m_b$ expansion in heavy quark effective theory (HQET) \cite{review}.
The $O(1/m_b^2)$ corrections have been computed in the
case of massless leptons
\cite{Bigi,bksv,MaWi}, and for the tau-lepton
in both the SM \cite{koy,bkps,us} and the 2HDM \cite{YZ,YZs}.
The one-loop QCD corrections to the tree-level amplitudes of the
spectator model are equally important.  In the SM,
the $O(\al)$ corrections have been computed for both
massless leptons \cite{CM,AP,CCM} and the
tau-lepton \cite{HoPa,CJK,ggsv}. However,
no such calculations are available for the 2HDM. Their absence
is a major source of uncertainty in the branching ratio
predictions for large $\tgb$ \cite{YZ}.

The $O(\al)$ corrections for $t \to bW$ \cite{JK}
and $t \to b H$ \cite{cd} are known.
The QCD corrections for the $\bctn$ decay can be directly obtained
from these results. This method was applied in \cite{CJK} to the
SM calculation. In this paper, we show that by working in the Landau
gauge, the method of \cite{CJK}
can be extended to include the charged Higgs contributions.

This paper is organized as follows.
In section II we calculate the $O(\al)$ corrections to the
$\bctn$ decay in the 2HDM (with model-II
Higgs-fermion couplings \cite{hhg}). In section III we
combine this new result with the known
$O(1/m_b^2)$ corrections and the quark mass relations of HQET,
and get an accurate prediction for the branching ratio
as function of $\tgb$ and the charged Higgs mass, $m_H$.
Section IV contains our summary.  Improvements
to the QCD computation through renormalization group resummation
are discussed in the Appendix.

\section{$\bold{O(\alpha_{\lowercase{s}}})$ Corrections to
$\bold{\bctn}$}

In \cite{CJK}, the QCD corrections to $\bctn$ in the SM
were obtained using the known QCD corrections of $t \to b W$
and $t \to b H$ decays, and without calculating any new Feynman diagrams.
The idea is to decompose the $n$ body phase space (where $n$=3 or 4)
into a product of $n-1$ body phase space and two-body phase space,
integrated over the invariant squared mass of the $\tau\nu$ pair
\cite{byckling}:
\beq \label{defpro}
d{\rm PS} (\bctn) \sim \int\,dq^2 \, d{\rm PS}(b \to X W^*)
\, d{\rm PS}(W^* \to \tau \nu),
\eeq
where $q^\mu = p^\mu_\tau + p^\mu_\nu$ is the lepton pair momentum,
$X$ is either $c$ or $c+$gluon, and $d$PS is the
appropriate phase space differential element.
The SM rate can be decomposed into two terms \cite{CJK}: one
corresponds to transverse $W$ exchange and the other is equivalent
to an (un-physical) scalar exchange.  The latter can be identified
with the exchange of a massless Goldstone boson in the Landau gauge.
When generalizing the analysis of \cite{CJK} to include a charged
Higgs boson, there are two additional terms: one that arises purely
from the charged Higgs exchange and the other that arises from the
interference between the $W$ and the charged Higgs diagrams.
The method proposed in \cite{CJK} is now easily extended
to this case by explicitly working in the Landau gauge. Then, there
are three diagrams contributing to $\bctn$ corresponding to
$W$-exchange, Goldstone boson exchange, and charged Higgs boson
exchange.  It is easy to show that the rate
can be decomposed into the sum of two incoherent decays:
\beq \label{gen}
\M = \M_W+\M_G+\M_H \qquad \Lra \qquad |\M|^2 = |\M_W|^2+|\M_G+\M_H|^2.
\eeq
The proof of (\ref{gen}) is straightforward.
The lepton tensor in interference terms
between a scalar (a Higgs or a Goldstone boson) and a vector boson ($W$)
carries one Lorentz index, $L_\mu$. After integration over the
$\tau\nu$ phase space, $L_\mu$ must be proportional to the only Lorentz
vector in the problem, $q_\mu$ (the lepton pair momentum).
Contracting $L_\mu$ with the Landau gauge $W$ propagator,
$T_{\alpha \mu} \sim g_{\alpha \mu} q^2 - q_\alpha q_\mu$, we get
\beq \label{cont}
L^\mu T_{\alpha \mu}\propto q^\mu(g_{\alpha \mu}q^2-q_\alpha q_\mu)=0\,.
\eeq
This proves (\ref{gen}). Note that this
proof holds for arbitrary $X$ in (\ref{defpro}), and is therefore
applicable to the QCD corrected rate to all orders in $\alpha_s$
(at lowest order in the electroweak coupling expansion).

Let us define
\beq \label{defvars}
\r\equiv{m_c^2 \over m_b^2},\ \ \ \e\equiv{m_\tau^2 \over m_b^2},\ \ \
t\equiv{q^2 \over m_b^2},\ \ \ \xi\equiv{m_H^2 \over m_b^2}.
\eeq
Due to (\ref{gen}) we can write schematically (in the Landau gauge)
\begin{eqnarray} \label{genint}
&&\G(\bctn)=\int_{\e}^{(1-\sqrt{\r})^2} {d\G \over dt} dt, \\
&&{d\G \over dt}=
H^{\alpha\beta}(b\to cW^*) T_{\alpha\mu}T_{\beta\nu}
L^{\mu\nu}(W^*\to \tau\nu)
 + H(b \to c S^*) L(S^* \to \tau \nu)\,, \nonumber
\end{eqnarray}
where $H$ and $L$ refer to the terms that arise from analyzing the
hadronic side and the leptonic side of the decay process, respectively.
As above, $T_{\alpha\mu}$ is the Landau gauge $W$ propagator.  The two
terms on the right hand side of the above equation correspond to the
sum over the two incoherent decays in (\ref{gen}).  In particular,
$S$ stands for an effective scalar that we construct to give the
contributions of both the Goldstone boson and the physical charged Higgs.
Moreover, it is convenient to absorb the Goldstone boson and charged
Higgs propagator factors ($-1/q^2$ for the
Goldstone boson and $1/m_H^2$ for the
physical Higgs boson in the approximation that $q^2 \ll m_H^2$)
directly into the definitions of the effective couplings
of $S$ to the leptons.  As a result, the scalar propagator factors
do not explicitly appear in (\ref{genint}).
The couplings of $S$ to quarks are then given by
\beq \label{defcoup}
{i g m_b V_{cb}\over 2 \sqrt{2} m_W}  (a + b \gamma_5),
\eeq
where $a\equiv a_G+a_H$, and $b\equiv b_G+b_H$,
with \cite{hhg}
\begin{eqnarray} \label{scalarcs}
a_G&=&-(1-\sqrt{\r}),\phantom{a_H=}\qquad\qquad
b_G=-(1+\sqrt{\r}),\phantom{b_H=} \nonumber \\
a_H&=&\tgb+\sqrt{\r}\ctgb,\qquad\qquad b_H=\tgb-\sqrt{\r}\ctgb.
\end{eqnarray}
The couplings of $S$ to leptons (with the propagator factors included
as indicated above) are given by
\beq \label{defcouplep}
{i g \over 2 \sqrt{2} m_W m_b}  (a^\ell + b^\ell \gamma_5).
\eeq
where $a^\ell\equiv a^\ell_G+a^\ell_H$, and
$b^\ell\equiv b^\ell_G+b^\ell_H$, with \cite{hhg}
\begin{eqnarray} \label{scallep}
a^\ell_G&=&{\sqrt{\e} \over t},\phantom{a_H=}\qquad\qquad
b_G^\ell={-\sqrt{\e} \over t},\phantom{b_H=} \nonumber\\
a_H^\ell&=&{\sqrt{\e}\tgb \over \xi},\qquad\qquad
b_H^\ell={-\sqrt{\e}\tgb \over \xi}.
\end{eqnarray}
Integrating over the leptonic two-body phase space yields $L$
[see (\ref{genint})]
\beq \label{lepten}
L(t)\sim  \le(1-{\e \over t}\ri)^2 t \e  \le[{1 \over t^2} -
{2\tgb \over t \, \xi} + {\tgbs \over \xi^2}\ri].
\eeq

We denote $\G_{abc}\equiv\G(a\to bc)$.
The expression for $\G_{tbH}$ can be found in \cite{cd}.
For arbitrary quark flavors and scalar Yukawa couplings,
Eqs. (7) and (8) in \cite{cd} can be parameterized as
$\G_{QqS}(m_q^2/m_Q^2,m_S^2/m_Q^2;c_1,c_2,c_3)$.
The relevant combinations of couplings are
\begin{eqnarray} \label{relterm}
c_1&=&a^2+b^2=
2(1+\r)+2(\tgbs+\r\ctgbs)-4(\tgb-\r\ctgb), \nonumber\\
c_2&=&a^2-b^2=
4\sqrt{\r}\le[-1 + \tgb\ctgb + (\tgb-\ctgb)\ri], \nonumber\\
c_3&=&ab=
(1-\r)+(\tgbs-\r\ctgbs)-2(\tgb+\r\ctgb).
\end{eqnarray}
Then, the contribution to the $\bctn$
decay rate from the effective scalar $S$ is obtained by taking
$S$ to be off-shell and integrating over its momentum:
\beq \label{finfur}
\G_S(\bctn)\sim\int_{\e}^{(1-\sqrt{\r})^2}
\G_{bcS}(\r,t;c_1,c_2,c_3)\, L(t)\, dt\,,
\eeq
where $L(t)$ is given in (\ref{lepten}) and $c_1$, $c_2$
and $c_3$ are given in (\ref{relterm}).

\section{Numerical Analysis}
In our numerical analysis we set $\ctgb=0$, as we are always interested
in large $\tgb$, and we
neglect the tiny $b \to u$ transitions. Since $\G_{QqS}$
is linear in the three relevant terms
in  (\ref{relterm}) we can identify the Higgs
and the interference terms in the total rate
\begin{eqnarray} \label{totr}
\G_H(\bctn)&=&{G_F m_b^2 \over 4\sqrt{2} \pi}\int_{\e}^{(1-\sqrt{\r})^2}
\G_{bcS}(\r,t;2,0,1)\,
\le(1-{\e \over t}\ri)^2 {t \e  \tan^4\beta \over \xi^2} \, dt\,,
\nonumber \\
\G_I(\bctn)&=&-{G_F m_b^2 \over 2\sqrt{2} \pi}\int_{\e}^{(1-\sqrt{\r})^2}
\G_{bcS}(\r,t;2,-2\sqrt{\r},1)\,
\le(1-{\e \over t}\ri)^2 {\e  \tgbs \over \xi} \, dt \,,
\end{eqnarray}
where $\G_H$ ($\G_I$) is the Higgs (interference) term in $\G$.
The tree level decay rate can be obtained by inserting the
appropriate tree-level expression for $\G_{bcS}$ into
(\ref{totr}).  We have performed the
integrals analytically and checked that we recover the known
results \cite{Kali,YZ}.   Turning to the $O(\alpha_s)$ computation,
we insert the appropriate one-loop expressions for $\G_{bcS}$ deduced
from \cite{cd}.  The integrals in (\ref{totr}) are then evaluated
numerically.

Our results are not very sensitive to small variations in the choice
of the $c$ and $b$ pole masses.
Below we quote results for representative values of the pole masses
$m_c=1.4$ GeV and $m_b=4.8$ GeV.
Decomposing the total rate into the tree-level and the one-loop parts,
\beq \label{dec}
\G=\G^0+\al\G^1,
\eeq
it is convenient to quote results for $\G^1/\G^0$ (note that factors
of $\tan\beta$ cancel in the ratio).  We get
\beq \label{num}
{\G^1_H \over \G^0_H}= -0.753\,,\ \ \
{\G^1_I \over \G^0_I}= -0.487\,.
\eeq
The SM corrections were calculated in \cite{us} based on
the calculations of \cite{HoPa}, and in \cite{CJK}:
\beq \label{numSM}
{\G^1_W \over \G^0_W}= -0.450 \,,\ \ \
{\G^1_e \over \G^0_e}= -0.545 \,,
\eeq
where $\G_W$ ($\G_e$) is the $W$ mediated rate for $\bctn$
($\bcen$).
Defining $\eta \equiv \G/\G^0$, and using $\alpha_s(m_b) \simeq 0.22$
[corresponding to $\alpha_s(m_Z)=0.115$], we get
\beq \label{finnum}
\eta_e \simeq 0.88, \ \ \
\eta_W \simeq 0.90, \ \ \
\eta_I \simeq 0.89, \ \ \
\eta_H \simeq 0.83.
\eeq

The main uncertainties in the calculation are from the unknown
$O(\al^2)$ corrections. To account for these, we vary
$\alpha_s$ in the range
$0.20 \leq \alpha_s(\mu) \leq 0.36$,
corresponding to $m_b/3 \leq \mu \leq m_b$ and
$0.110 \leq \alpha_s(m_Z) \leq 0.125$.
This leads to the ranges
\beq \label{finrater}
1.02 \leq {\eta_W \over \eta_e} \leq 1.04, \ \ \
1.01 \leq {\eta_I \over \eta_e} \leq 1.03, \ \ \
0.96 \geq {\eta_H \over \eta_e} \geq 0.91.
\eeq
Note that these ratios are rather insensitive to the exact choice of
$\alpha_s$.  This completes the analysis of the $O(\alpha_s)$
corrections.

To arrive at our final prediction, one must also include
the $O(1/m_b^2)$ corrections to the spectator model.
In the Heavy Quark Expansion, we use the HQET mass relation
\begin{equation}\label{cbmasses}
m_B = m_b+\bar\Lambda-{\lambda_1+3\lambda_2\over2m_b}+\ldots\,,\qquad
m_D = m_c+\bar\Lambda-{\lambda_1+3\lambda_2\over2m_c}+\ldots\,,
\end{equation}
and the numerical values as in \cite{YZ}
\begin{equation}\label{ranges}
0.4< \bar\Lambda <0.6\,{\rm GeV}\,,\ \ \
0< -\lambda_1 <0.3\,{\rm GeV}^2 \,, \ \ \
0.11< \lambda_2(m_b) < 0.13\,{\rm GeV}^2 \,.
\end{equation}
Combining the $O(1/m_b^2)$ calculation of \cite{YZ}
with our $O(\al)$ calculation, and using the
range of the various parameters as given in
Eqs. (\ref{finrater}) and (\ref{ranges}), we
get the theoretical prediction for the branching ratio (normalized
to the electron channel) as a function of
$r\equiv \tgb/m_H$ given in Fig.~1. Comparing it with Fig.~1(a)
of \cite{YZ} we learn that, as anticipated in \cite{YZ},
the errors in the plot are reduced by about
a factor of two for large $r$.  The residual errors
arise almost entirely from the unknown
pole masses, namely the values of $\bar \Lambda$ and $\lambda_1$.

Combining the recent measurements of the branching ratio of $\bctn$
from ALEPH \cite{Aleph} and L3 \cite {L3} and using the world
average for $\bcln$ ($\ell =e,\mu$) \cite{PDG},
\begin{equation}
{\rm BR}\,(\bctn) = 2.69\pm 0.44\,\%\,, \qquad
{\rm BR}\,(\bcln) = 10.43\pm0.24 \,\%\,,
\end{equation}
we obtain the $2\sigma$ upper bound
\beq \label{bound}
r < 0.52\,{\rm GeV}^{-1}\,.
\eeq
The $1\sigma$ upper bound is $r<0.49\,{\rm GeV}^{-1}$
(as compared to $r<0.51\,{\rm GeV}^{-1}$ in \cite{YZ}).
The improvement over the bounds obtained in \cite{YZ}
originates from both the experimental improvement
and from our $O(\al)$ calculation.

Our analysis holds with minor modifications for general multi-Higgs
doublet models with natural flavor conservation
(for a recent analysis, see \cite{Yuval}).
In such models, instead of the single parameter $\tan\beta$, three
complex coupling constants determine the Yukawa interactions of the
lightest charged scalar (assuming that the heavier charged scalars
effectively decouple from the fermions).
The parameters $X$, $Y$ and $Z$ describe the couplings
to down-type quarks, up-type quarks and charged leptons, respectively.
The bounds on $\sqrt{|XZ^*|}/m_H$ and
$\sqrt{|{\rm Re}(XZ^*)|}/m_H$ are the same as that on $r$.
The $2\sigma$ upper bound on the imaginary part of $XZ^*$ is
$\sqrt{|{\rm Im}(XZ^*)|}/m_H < 0.44\,{\rm GeV}^{-1}$
($<0.41\,{\rm GeV}^{-1}$ at $1\sigma$).

\section{Summary}
We have calculated the $O(\al)$ corrections to $\bctn$ decay in the 2HDM.
In addition to the known SM result, we have computed the
contributions arising from the exchange of a charged Higgs boson;
both the pure Higgs term and the $W^+$--$H^+$ interference term
have been included.  Working in the Landau gauge,
the calculation is carried out by making use of previous calculations
of the QCD corrections to $t \to b H$.
Our results for the numerical value of the QCD corrections
are given in Eq.(\ref{finnum}).
We combined our calculation with the known
$O(1/m_b^2)$ corrections and the quarks mass relations of HQET,
and obtained an accurate prediction for the branching ratio
as function of $\tgb/m_H$, shown in Fig.~1.  The
remaining uncertainties in the prediction are almost entirely from
the uncertainties in the pole masses.

The virtual effects of the charged
Higgs boson can be important in other processes.
However, in most cases, the charged Higgs exchange
only first enters at the one-loop level.  In models with
an extended Higgs sector, there is often additional new physics
beyond the Standard Model that can also contribute to the
processes under consideration.  In this case, there can be no clean
bound on Higgs sector parameters, since the one-loop charged Higgs
effects can be partially canceled by the effects of
other particles in the theory. For example, in the minimal
supersymmetric model, the bounds on the charged Higgs parameters
from the one-loop charged Higgs contributions to
$b \to s \gamma$ \cite{bsg} and $Z \to b \bar{b}$ \cite{zbb}
decays can be avoided if partially canceled
by loop diagrams with intermediate superparticles.
In contrast, the charged Higgs boson
contribution to $\bctn$ studied in this paper occurs
at tree level; therefore, its effect cannot be
canceled by other new particles in the theory.

\acknowledgments
This work was supported in part by the United States -- Israel
Binational Science Foundation (BSF).
YG thanks Zoltan Ligeti for useful discussions.
HEH would like to gratefully acknowledge the kind hospitality
of the Weizmann Institute of Science, where this work was completed.
The work of YN is supported in part by the BSF, by the
Israel Commission for Basic Research and by the Minerva Foundation.

\appendix
\section{Renormalization Group improvement}

The QCD corrections to $\G(t \to b H)$ were
calculated for $m_b=0$ in \cite{LiYuan,LiuYao},
and extended for finite $m_b$ in \cite{cd,csl}.  Hence,
we use the results of \cite{cd}\footnote{The results of \cite{csl}
differ from those of \cite{cd}.  However, the latter results appear
more reliable; see the discussion of \cite{cd} for further
clarifications.}.
All calculations were performed in the on-shell
mass renormalization scheme, in which the quark
pole masses appear in the formulae.
These results can be improved by using the technology of the
renormalization group equations (RGE) to resum the leading logarithms
to all orders in $\alpha_s$.  In particular, note that
for small $\epsilon \equiv m_b^2/m_t^2$, $\al\log\epsilon$ is
typically of O(1).  Consequently, all leading logarithmic terms of
the form  $\epsilon\,\al^n \log ^n \epsilon$ should be summed
\cite{BL}.  This can be accomplished by noting that
one of the effects of the one-loop QCD corrections
is the replacement of the pole mass $m_b$ with the one-loop running mass
$m_b(m_t)$ \cite{yossi}.  By replacing the one-loop
$m_b(m_t)$ with the fully RGE-integrated running mass,
$m_b^{RGE}(m_t)$, one successfully
resums all $\alpha_s\log\epsilon$ terms.

The one-loop QCD corrections to $\G(t\to bH)$ were obtained in
\cite{cd}, but the RGE-improved rate was not given.  To carry out the
RGE-improvement, first expand the results of \cite{cd} in powers of
$\epsilon$.  Formally, the full QCD-corrected rate can be written as
\beq \label{pertseries}
\G=\sum_{nm}\,\epsilon^n\alpha_s^m\G_n^m\,,
\eeq
which defines $\G_n^m$.  In this case, only terms up to $O(\alpha_s)$
are known.  It is sufficient to keep
terms up to $O(\epsilon)$ since $\epsilon\simeq 10^{-3}$
is a small number, and only
the $O(\epsilon)$ term can be enhanced by the possibly
large $\tgbs$ coefficient.  Thus, we write
$\G=\G_0^0+\al\G^1_0 + \epsilon\le(\G^0_1+\al\G_1^1\ri)$.
Using the expressions given in \cite{cd}, we find
\beq \label{epsexp}
\G=\G_0^0+\epsilon\G^0_1\le(1+{4 \al \over \pi} \log\epsilon \ri)
+ \al\G^1_{fin},
\eeq
where $\G^1_{fin}$ contains no $\log\epsilon$ terms, and
\begin{eqnarray} \label{exaeps}
\Gamma_0^0&=&{G_Fm_t^3|V_{tb}|^2\over8\sqrt2\pi}(1-\chi)\cot^2\beta\,,
\nonumber \\
\Gamma_1^0&=&{G_Fm_t^3|V_{tb}|^2\over2\sqrt2\pi}
\left\{(1-\chi)\left[1+{1\over4}(1-\chi)\tan^2\beta\right]
-{\chi\over2}\cot^2\beta\right\}\,,
\end{eqnarray}
where $\chi \equiv m_H^2/m_t^2$.  Using
\beq \label{mbmt}
m_b^2(m_t) = m_b^2(m_b) \le(1+{4 \al \over \pi} \log\epsilon +
O(\al^2) \ri),
\eeq
we see that (\ref{epsexp}) is equivalent at $O(\al)$
to replacing $m_b(m_b)$ with $m_b(m_t)$ in $\G_0^0+\epsilon\G^0_1$.
This is a general result \cite{yossi}. Then, to sum up the
$\epsilon\,\al^n \log ^n \epsilon$ terms, we replace the
$O(\al)$ expression for $m_b(m_t)$ with $m_b^{RGE}(m_t)$ \cite{BL}:
\beq \label{replace}
m_b(m_t) = m_b(m_b) \le(1+{2 \al \over \pi} \log\epsilon \ri)
\ \Longrightarrow \
m_b^{RGE}(m_t) = m_b(m_b)
\le[{\al(m_b) \over \al(m_t)}\ri]^{-2 \gamma_0/\beta_0},
\eeq
where $\gamma_0=2$ and $\beta_0=11-2/3 n_f$.

The decay rate depends on $m_b$ through both the Yukawa coupling
and the phase space integrals. Far from threshold, the dependence of
the phase space integrals on $m_b$ is weak; therefore, the
RGE-improvement has no numerical significance there.
In contrast, the RGE-improvement via the
Yukawa couplings is important provided that $\tgb$ is large
(it is only for large $\tgb$ that the rate is sensitive to $m_b$).
To see the effect, we define
$\eta \equiv \G/\G^0$ prior to RGE-improvement (see section III),
where in the notation above, $\G^0\equiv\sum_n\epsilon^n\G_n^0$.
The results of the RGE-improvement are sensitive
to the procedure that we take.  We can use the RGE-improved $m_b(m_t)$
in $\G^0$ only (we call the result $\eta^{RGE}_0$)
or in both $\G^0$ and $\G^1_{fin}$ ($\eta^{RGE}_1$). The difference
is $O(\alpha_s^2)$, beyond the accuracy of the calculation, but
in practice it can turn out to be important.
We take $m_b=4.8$ GeV, $m_t=175$ GeV, $m_H=100$ GeV and
$\alpha_s(m_t)=0.1$. As a first example, take $\tgb=1$.
Then $\eta\sim\eta^{RGE}_0\sim\eta^{RGE}_1\sim 0.92$, with  difference
of $O(10^{-4})$.   We see that the weak dependence
of the rate on $m_b$ makes the RGE-improvement unimportant.
As a second example, take $\tgb=100$:
\beq \label{etalargetgb}
\eta \sim 0.46,\ \ \
\eta^{RGE}_0\sim 0.50,\ \ \
\eta^{RGE}_1\sim 0.46\,.
\eeq
The difference between $\eta^{RGE}_0$ and $\eta^{RGE}_1$ is a measure
of the size of the neglected $O(\alpha_s^2)$ terms.   Given that
$\eta\simeq\eta^{RGE}_1$, it follows that for very large $\tan\beta$,
the RGE-resummed result does not provide a practical improvement over
the original one-loop calculation.

The situation is different for the $\bctn$ decay in three important ways:

(i) The RGE improvement is related to
the Yukawa coupling of the light quark. In $\bctn$,
the Yukawa coupling of the charm can be safely neglected.

(ii) The RGE improvement is important to sum
large logs. However, $\log (m_b/m_c) \sim O(1)$ is not large.

(iii) The RGE improvement is a correction to the $O(\al)$
corrections, but as noted below (\ref{finrater}), the latter
almost cancel in the relevant ratio of rates that we consider.

Therefore, while theoretically important, the RGE improved result
has no numerical significance for the $\bctn$ calculation.

{\tighten

\begin{figure}
\epsfysize=14truecm
\centerline{\epsfbox{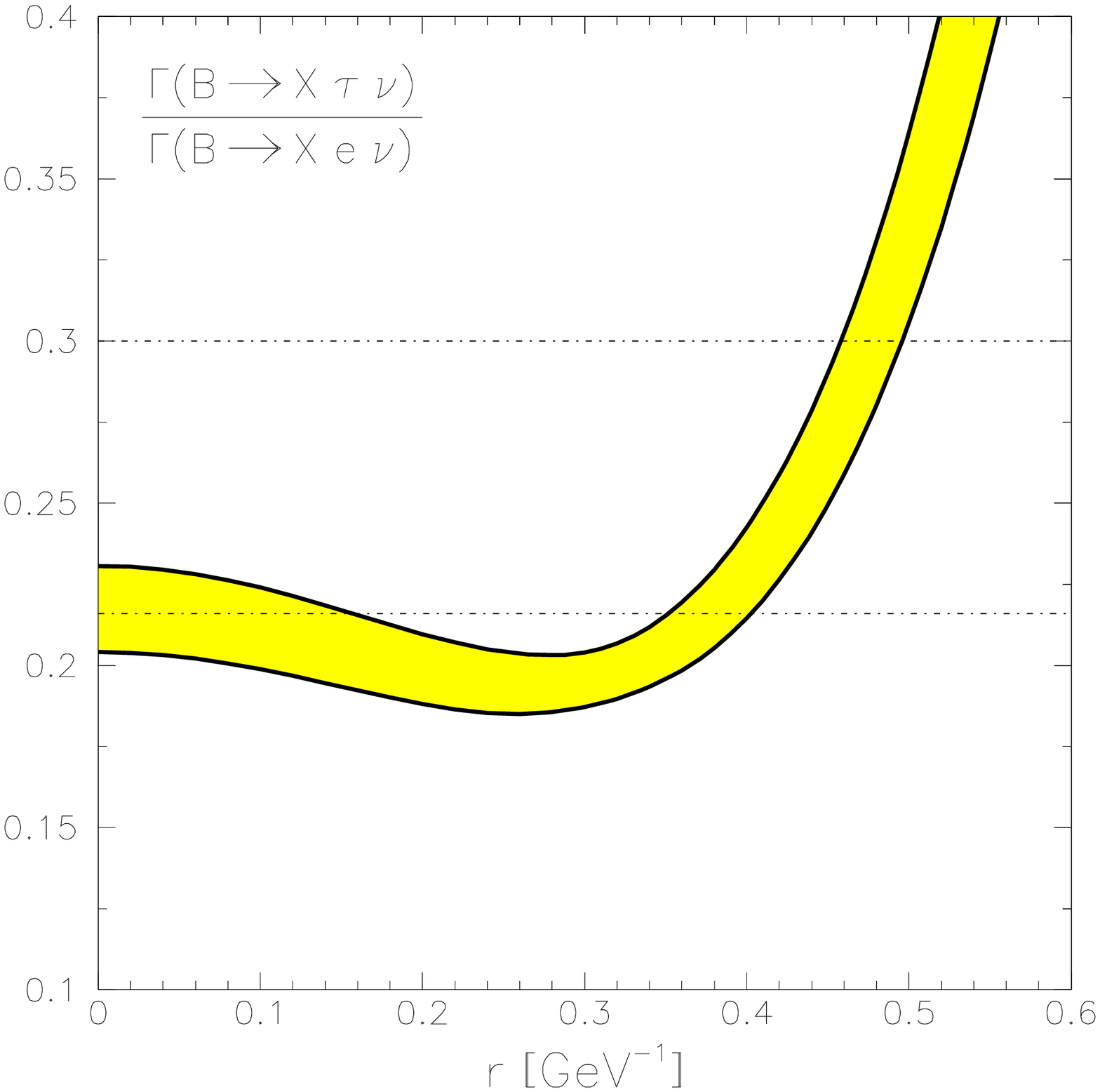}}
\vspace{7mm}
\caption[a]{
$\Gamma(\bctn)/\Gamma(\bcen)$
as a function of $r=\tan\beta/m_H$.  The shaded area between the
solid lines is our result. The dash-dotted lines give the
experimental $1\sigma$ bounds.  }

\end{figure}
}
\end{document}